\documentclass{aa}
\usepackage{graphicx,amssymb}
\newcommand{\mode}{$\langle I\rangle$}
  
\def\as{\arcsec\hspace{-1.2mm}.\hspace{0.3mm}} 
\def\am{\arcmin\hspace{-1.2mm}.\hspace{0.3mm}}  
\def\e{e$^-$}

\begin{document}
   \title{A robust algorithm for sky background computation in CCD images
   \thanks{Partially based on observations collected at ESO-Paranal.}}

   \author{F. Patat}
   \offprints{F. Patat;  \email{fpatat@eso.org}}

   \institute{European Southern Observatory, K.Schwarzschild Str.2,
              85748 - Garching - Germany\\}

   \date{Received October 24, 2002; accepted Januar 24, 2003}

\abstract{In this paper we present a non-interactive algorithm to estimate 
a representative value for the sky background on CCD images. The 
method we have 
devised uses the mode as a robust estimator of the background brightness in 
sub-windows distributed across the input frame. The presence of contaminating 
objects is detected through the study of the local intensity distribution 
function and the perturbed areas are rejected using a statistical criterion 
which was derived from numerical simulations.
The technique has been extensively tested on a large amount of
images and it is suitable for fully automatic processing of large data
volumes. The implementation we discuss here has been optimized for the
ESO-FORS1 instrument, but it can be easily generalized to all CCD imagers
with a sufficiently large field of view. The algorithm has been successfully 
used for the $UBVRI$ ESO-Paranal night sky brightness survey (Patat 
\cite{patat}).

}

\titlerunning{Sky background computation}
\authorrunning{Ferdinando Patat}

\maketitle

%
%________________________________________________________________

\section{\label{sec:intro} Introduction}

The study of the night sky brightness is fundamental to monitor
the quality of very dark astronomical sites and sets the stage
for the study of a whole series of effects which take place in the
upper layers of Earth's atmosphere (Leinert et al. \cite{leinert98} 
and references therein). 

With the exception of a very few cases, all night sky brightness surveys
are executed with photoelectric devices coupled to small telescopes and
usually span a limited number of nights (Benn \& Ellison \cite{benn}), 
distributed across several years. For these reasons, the data are usually 
rather scanty and suffer from the inclusion of bright stars 
($V\geq$13; see for example Walker \cite{walker88}). 

Nowadays, with the availability of large telescopes equipped with CCD
imagers and the possibility of reducing the data via dedicated
pipelines, the paradigm is changing and different approaches become feasible.

In this spirit, during the beginning of year 2000 we have started a
project to monitor the $UBVRI$ night sky brightness at ESO-Paranal 
Observatory (Chile) as part of the quality control (QC) 
procedures implemented for the FOcal Reducer/low dispersion Spectrograph 
(hereafter FORS1). This multi-mode optical instrument, which is mounted at the
Cassegrain focus of ESO-Antu/Melipal 8.2m telescopes (Szeifert 
\cite{szeifert}), has two remotely exchangeable collimators, which give an 
imaging field of view of 6\am8$\times$6\am8 (standard resolution, SR) and 
3\am4$\times$3\am4 (high resolution, HR) respectively.  

FORS1 is offered during dark time both in Visitor Mode (VM) and Service Mode 
(SM). Imaging data obtained during SM runs are bias and flat-field corrected
by the pipeline and undergo a series of quality checks before they are finally
distributed to the users. Due to the high number of 
imaging frames produced by this instrument (more than 4500 from April 
2000 to September 2001) and the variety of scientific cases which drive it, 
it is clear that a complete and systematic study of the night sky brightness 
can be performed only by means of a robust and automatic procedure, capable 
of identifying and rejecting all the cases which are not suitable for sky 
background measurements (e.g. large galaxies, crowded stellar fields and so 
on). 

In this work we present and discuss the algorithm we have specifically
designed for this purpose, while the results of the night sky brightness 
survey are reported in Patat (\cite{patat}).

The paper is organized as follows. In Sec.~\ref{sec:bgr} we discuss the
problems connected with the sky background measurement in digital images,
while Sec.~\ref{sec:mode} deals with the technique we have adopted
to compute the mode of the image intensity distribution. The algorithm we 
have devised to identify the presence of contaminating objects in the field
and the tests we have performed on real FORS1 data are presented in 
Sec.~\ref{sec:delta} and \ref{sec:test} respectively.
Finally, in Sec.~\ref{sec:conclusion} we summarize our conclusions.

\section{\label{sec:bgr} Problems in estimating the sky background}

Widely available programs for object detection and photometry like
DAOPHOT (Stetson 1987) and Sextractor (Bertin \& Arnouts 1996) use
the mode of the image intensity distribution to estimate the local
background or to construct the background
map of a two-dimensional frame. In fact, besides its robustness, the mode 
is a statistically powerful estimator, being the most probable value of
the background brightness in a given region of the image.

The use of the mode as the maximum-likelihood estimator
for the {\it sky} background implicitly assumes that the image has some 
regions which are not seriously contaminated by any astronomical object. Of 
course distant and faint galaxies (and stars) will always be present, but we 
will consider them as part of the global background throughout this paper. 
Therefore, when we talk about contaminating objects, we mean objects which 
are detectable in the image, well above the background noise. There is also 
another assumption that one has to make, i.e. that it is really improbable
that the images to be analysed are filled by a single extended 
object with a roughly constant surface brightness. In fact this is the only 
case where one would have an overall background increase without having any 
secondary effects on the intensity distribution function and in that case the 
process would lead to an overestimate of the sky background. In the case of
FORS1 frames, especially with the commonly used SR collimator field of
6\am8$\times$6\am8, this assumption seems reasonable.

Due to the field of view of FORS1 and the wide variety of scientific 
projects which are carried out by this multi-mode instrument, one 
expects to deal with very different astronomical objects, which
would perturb in a different way the local sky background. Possible examples
are comets, clusters of galaxies, outskirts of big spirals, large ellipticals,
diffuse nebulosities, crowded stellar fields and so on. In all these cases 
there still might exist 
parts of the image which are suitable for a sky background measurement. For 
this reason it is clear that the analysis has to be performed using 
sub-windows distributed on a grid across the input image. The choice of the 
sub-window size has to be done in such a way that this is neither too small, 
because in that case the fraction of uncontaminated pixels might become 
statistically insignificant, nor too big, 
otherwise the probability  of including large diffuse objects becomes
large. After running some tests we have seen that a 300$\times$300 px 
sub-window (which corresponds to 1 arcmin$^2$ for the SR collimator) gives
satisfactory results. Since the guide probe of FORS1 is sometimes vignetting 
the outer parts of the images, we have decided to use the central 
1800$\times$1800 px region of the detector only. This allows one to analyse 
the images in a 6$\times$6 sub-windows grid including 9$\times$10$^4$ px each.
 
Of course there is always the possibility that none of the sub-windows is 
{\it clean} enough to allow for a reliable measurement. 
A few examples are galactic stellar fields with heavily saturated stars, 
outer parts of globular clusters, nearby interacting galaxies and close-by 
comets, just to cite a few real examples we encountered during this analysis.

\begin{figure}
\resizebox{\hsize}{!}{\includegraphics{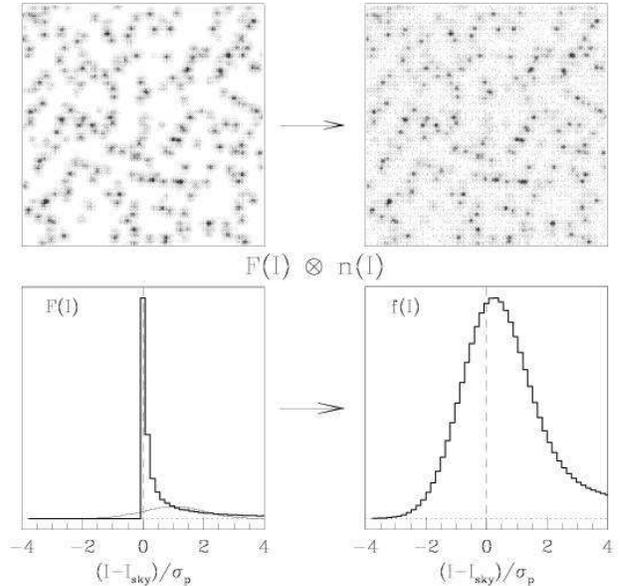}}
\caption{\label{fig:degrad} Illustration of the effect on the intensity
distribution when a Poissonian noise is added to a stellar field with a
constant background $I_{sky}$. The vertical scale in the lower plots is 
arbitrary, while the intensity $I$ is counted from the input $I_{sky}$ 
value and normalised to $\sigma_p=\sqrt{I_{sky}}$. The solid thin curve in the
lower left plot depicts the contribution to the final intensity distribution
by the value of $F(I)$ at $I=I_{sky}+\sigma_p$.
}
\end{figure}

\begin{figure}
\resizebox{\hsize}{!}{\includegraphics{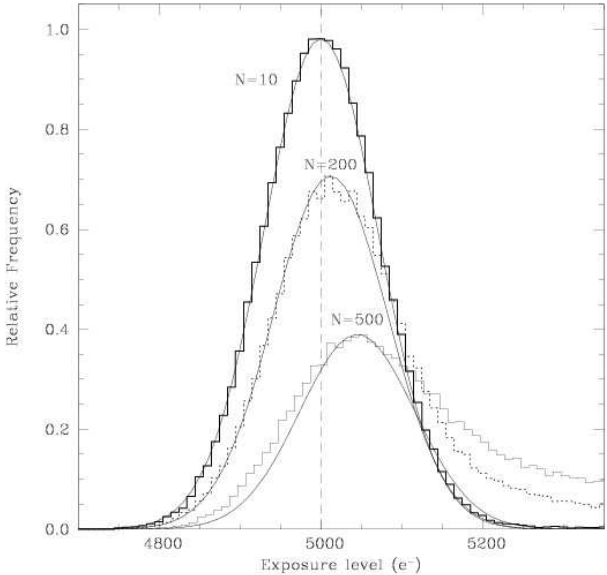}}
\caption{\label{fig:simul} Intensity distributions for three simulated images
obtained injecting 10, 200 and 500 stars (from top to bottom) on a 
background $I_{sky}$=5000 \e\/ with a Poissonian noise distribution. The
solid curves are Gaussian profiles centred on the distribution mode
and having the $\sigma$ expected for a Poissonian noise (see text
for more details).}
\end{figure}

The difficulties one faces in estimating the background are easily
understood from the following considerations. If the images to be analysed
were noiseless and the sky background constant and equal to $I_{sky}$, 
the solution of the problem would be trivial. In fact, in this case, 
the corresponding intensity distribution function $F(I)$ would be 0 for 
$I<I_{sky}$, would then suddenly peak at $I=I_{sky}$ and finally show an
extended tail, whose shape depends on the number and intensity of 
contaminating objects. This is illustrated
in the left part of Fig.~\ref{fig:degrad}, where we have presented an
artificial stellar field and its intensity distribution. As usual, Nature
behaves in a more subtle way and due to the photon statistics (and
marginally to detector read-out) the images we are going to deal with
are always affected by noise.
From a mathematical point of view it is very easy to predict the noise
effect on the intensity distribution. In fact, if $n(I)$ is the
noise distribution, the observed intensity distribution is simply
given by:

\begin{equation}
\label{eq:degrad}
f(I)=\int_{-\infty}^{I} F(I-z) \; n(I,z) \; dz
\end{equation}

i.e. the convolution $F\otimes n$ of the signal $F(I)$ 
with a variable response function $n(I,z)$, which can be expressed as
$n(I,z)=1/\sqrt{2\pi\;z} \;\; exp[-\frac{1}{2}(z^2/(I-z)]$, provided that
the read-out noise is negligible. From these
simple considerations it is clear that what really matters for the 
degradation of the resulting peak sharpness and the consequent uncertainty 
in the estimate of its original position is the shape of the input 
distribution function $F(I)$ within $\sim$5$\sqrt{I_{sky}}$ from its peak, 
while the behaviour at higher intensities is unrelevant (see 
Fig.~\ref{fig:degrad}, right plots).
With respect to these effects, it is quite instructive to look at the results 
of some numerical simulations. In Fig.~\ref{fig:simul} we have plotted the
relative intensity distributions of three 300$\times$300 px artificial
frames with a fixed value of the sky background ($I_{sky}$=5000 electrons)
on which we have randomly injected 10, 200 and 500 stars respectively.
We have used a Moffat profile for the stars with $\beta$=4 and a FWHM of
5.3 px. In the case of FORS1 and SR collimator this would correspond to
a field of 1 arcmin$^2$ and a seeing of 1$^{\prime\prime}$.
The peak intensity of the stars was randomly generated in the
range 0-10$^4$ electrons. 

Basically three effects are visible as the number of stars grows: A) the mode 
\mode\/ of the distribution steadily increases; B) the distribution core 
width increases; 
C) the distribution becomes more and more skewed, with a tail appearing at 
the highest intensities. Clearly effect A and B will respectively lead 
to overestimates of the sky background $I_{sky}$ and the noise, which
in the case of an uncontaminated field is expected to be 
$\sigma_p=\sqrt{I_{sky}}$,
according to the Poissonian statistics. One possible solution to this problem
is achieved by reconstructing the intensity distribution one would have
if no contamination effects were present. A good example of such a solution
is represented by the asymmetric clipping algorithm developed by Ratnatunga 
\& Newell (\cite{ratna}), which computes the background distribution via an 
iterative clipping of the right wing of the perturbed intensity distribution.

Due to the high number of available FORS1 frames and the purpose of this work,
we can afford a different approach. Instead of attempting to reconstruct the 
underlying sky background intensity distribution in contaminated regions, we 
rather try to identify these regions and exclude them from all further 
calculations. 
For this purpose we have devised a simple and robust test that can be 
used to estimate the degree of contamination in an image and operated in
an automatic way on large amounts of data.

The first step in the application of our method is the mode estimate, which 
we describe in the next section.

\section{\label{sec:mode}Mode estimate}

The mode \mode\/ of a distribution $f(I)$ is defined as the 
most probable value of $I$ (see for example Lupton 1993) or, in other words, 
the value of $I$ where $f(I)$ takes its maximum value. For moderately skewed
distributions the mode can be approximately computed as $\langle I\rangle\simeq
3\times median - 2\times mean$ (Kendall \& Stuart 1997). Unfortunately, the
observed intensity distributions in FORS1 images often show very extended 
tails, due to several effects like saturated stars, cosmic-ray events and so 
on. While this tail usually does not affect the mode, it does perturb the 
mean and the above formula would lead to wrong results. A possible solution
is given by the application of this approximation after an iterative
clipping of the distribution around its median, as it is done for example
in Sextractor (Bertin \& Arnouts 1996). In practice one is forced to use
this method when one has to compute the background map of an image using 
small sub-windows. In fact in that case the statistics would be too poor 
to compute the mode in a reliable way directly using the distribution shape.

This is not the case here, since we are more interested in an average value
rather than in a map of the background within the same image. For this reason 
we can perform our analysis in much larger sub-windows so that we 
can build up a very good signal-to-noise distribution function. This allows 
us to compute the mode just using its definition, without any loss of 
generality and in a very robust way.

\subsection{\label{sec:obt} Introducing the Optimal Binning Technique (OBT)}

Since we are dealing with discrete distribution functions, finding the mode 
implies that the data have to be binned and the maximum of the distribution 
has to be found. 
To reduce the noise due to the finite (and possibly large) bin size $\Delta I$, 
we have chosen to refine the direct mode estimate using a quadratic 
interpolation on the modal bin and the two adjacent bins (see for instance 
Spiegel 1988). 
If $I_m$ and $f_m=f(I_m)$ are the values of the intensity and intensity 
distribution in the modal bin and $(I_r,f_r)$,$(I_l,f_l)$ are the 
corresponding values in the two adjacent bins respectively, then the 
mode \mode\/ is estimated as the value of $I$ where the interpolating 
parabola reaches its maximum. One can show that this is given by the 
following expression:

\begin{equation}
\label{eq:parabola}
\langle I\rangle = I_m + \frac{f_l-f_r}{f_l-2f_m+f_r} \; \frac{\Delta I}{2}
\end{equation}

This correction has the effect of moving the estimated mode 
$\langle I \rangle$ within the whole bin $\Delta I$ according to the local 
shape of the distribution function. The modal bin boundaries are reached when 
$f_m=f_l$ or $f_m=f_r$ (provided that $f_l \neq f_r$). Numerical 
simulations have shown that the parabolic interpolation is very 
effective in reducing the error on the mode estimate (see below), 
at least for the kind of distributions considered here. For more 
general cases, the interpolation option will improve the mode accuracy
only if the asymmetry of the distribution around the mode perturbs
$(f_l-f_r)$ significantly less than the average value of the correction
corresponding to an offset of the central bin.

The choice of bin size is critical. If the bin $\Delta I$ is too small the 
resulting histogram will be too noisy to give a robust estimate of \mode. 
On the other hand, if $\Delta I$ is too large, then the mode estimate will be 
affected by a large uncertainty due to the small signal-to-noise in the
two outer bins. For this reason we need to find an optimal value for 
$\Delta I$ which minimises the error on the mode. 

For this purpose we have run a series of numerical simulations
of uncontaminated windows with a known sky level on top of 
which we have added a Poissonian noise and a typical read-out
noise (6 electrons). For each of the simulated frames the mode was
estimated using the parabolic interpolation described above.
This has been done for different values of the 
bin size $\Delta I$ and the number of pixels $N=n_{pix}^2$ included 
in each testing window. For each pair ($\Delta I$,$n_{pix}$) we have
performed 5000 simulations. What one sees is that for a given value
of $n_{pix}$, the RMS deviation of the estimated mode from the known 
input value $I_{sky}$ decreases as $\Delta I$ increases, reaches a minimum 
and then grows again, as expected from the above considerations
(see also Fig.~\ref{fig:uncont}). 
For this reason it makes sense to adopt the value of $\Delta I$ where
the RMS error $\epsilon_{RMS}$ reaches its minimum as the optimum bin size 
for the mode estimate.

\begin{figure}
\resizebox{\hsize}{!}{\includegraphics{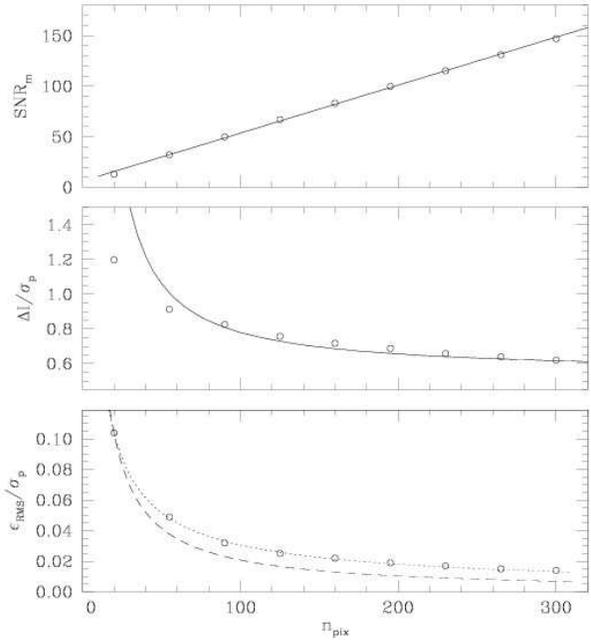}}
\caption{\label{fig:optfit}
Upper panel: Signal-to-noise ratio $SNR_m$ reached in the modal bin
for the optimal value of $\Delta I$ as a function of $n_{pix}$. The
solid line represents a linear fit to the data with $n_{pix}\geq$90.
Middle panel: normalised optimal bin width. The solid line traces
Eq.~\ref{eq:binsize}. Lower panel: optimal values of the RMS error on the mode.
The dotted line is a best fit to the data 
($\epsilon_{RMS}\propto n_{pix}^{-0.75}$) 
while the dashed line traces the law $\epsilon_{RMS}\propto n_{pix}^{-1}$. 
In all panels each point is the result of 5000 simulations.}
\end{figure}

Before we proceed with the presentation of the results, we want to
discuss a few more points. The pixel counts in the intensity
distribution bins obey Poissonian statistics. This implies
that if $f(I)$ is the number of pixels falling in a given bin, the
RMS uncertainty on $f(I)$ is simply given by $\sqrt{f(I)}$ and hence
we can define a signal-to-noise ratio $SNR=\sqrt{f(I)}$. In particular,
we can introduce $SNR_m$, i.e. the signal-to-noise ratio reached in the
modal bin. If $f$ is the fraction of pixels falling in the modal bin, we
have obviously 

\begin{equation}
\label{eq:fraction}
f=(SNR_m/n_{pix})^2
\end{equation}

Having this in mind, we can now have a look at the behaviour of $SNR_m$
as a function of $n_{pix}$ when the optimal bin size is used to estimate
the mode in our simulations. This is portrayed in the upper panel of
Fig.~\ref{fig:optfit}, where one can easily see that $SNR_m$ depends almost
linearly on $n_{pix}$. In the range $90\leq n_{pix}\leq 300$, the best fit
to the simulated data (see Fig.~\ref{fig:optfit}, upper panel) gives

\begin{equation} 
\label{eq:snrm}
SNR_m\simeq 6.2+0.48\;n_{pix}
\end{equation}

For large values of $n_{pix}$ this relation can be approximated as 
$SNR_m\approx n_{pix}/2$, which means that optimal results are obtained when 
$\sim$25\% of the pixels fall in the modal bin.

Having this result, it is easy to compute the optimal fraction $f$ of pixels
by means of Eq.~\ref{eq:fraction} and Eq.~\ref{eq:snrm} (or its approximate
expression). Then, assuming the underlying distribution $f(I)$ to be a 
Gaussian with $\sigma=\sigma_p$, one can derive the corresponding bin size 
in terms of $\sigma_p$ as 

\begin{equation}
\label{eq:di}
\Delta I=2\;k(f)\;\sigma_p
\end{equation}

where $k$ is implicitly defined by the following equation:

\begin{equation}
\label{eq:kappa}
f=\frac{1}{\sqrt{2\pi}\sigma_p} 
\int_{\langle I\rangle -k\sigma_p}^{\langle I\rangle +k\sigma_p}
\exp \left [ {-\frac{(I-\langle I \rangle)^2}{2\sigma_p^2}}\right ]\; 
dI
\end{equation}

This equation can be solved numerically and the solution fitted by
a low order polynomial. For $f\leq$0.33 ($n_{pix}>$60), the solution can
be approximated rather accurately by the following expression:

\begin{equation}
\label{eq:kf}
k(f)\simeq 1.28\;f
\end{equation}

Finally, using Eqs.~\ref{eq:di} and \ref{eq:kf}
one can easily compute the optimal bin width. For $n_{pix}\geq 90$ we can 
approximate the expression for the optimal bin as follows:

\begin{equation}
\label{eq:binsize}
\frac{\Delta I}{\sigma_p}\simeq 2.6 \; 
\left ( 0.48 + \frac{6.2}{n_{pix}} \right )^2
\end{equation}

while, for very large values of $n_{pix}$, the ratio between the
optimal bin size $\Delta I$ and $\sigma_p$ approaches asymptotically 
the value 0.6. The result is shown in the 
central panel of Fig.~\ref{fig:optfit}, where we have normalised the optimal
bin to $\sigma_p$ to remove the dependency on the sky level. The comparison
between the predicted values (solid line) and the results of the simulations
are in good agreement across all the $n_{pix}$ explored range.

Finally, the RMS deviation from the input value in our simulations clearly
decreases for increasing values of $n_{pix}$, as shown in the
lower panel of Fig.~\ref{fig:optfit}. For comparison we have overplotted
the $n_{pix}^{-1}$ law (dashed line) expected if the error would
scale proportionally to the overall signal-to-noise ratio. It is interesting 
to note that the RMS error in the simulations decreases at a slower rate,
approximately as $n_{pix}^{-0.75}$ (dotted line). We will come back to this 
point in the next section, when discussing the efficiency of the method in 
reducing the error; for the time being, we only notice that with $n_{pix}$=20
the expected RMS error for $I_{sky}$=100 electrons is about 1\%, which 
reduces to 0.2\% for $n_{pix}$=300.

\subsection{\label{sec:realcase} Application of OBT to the real case}

As we have discussed in Sec.~\ref{sec:bgr}, the real data show a variety of
contaminating effects, which tend to skew the intensity distribution and
affect in different ways the mode estimate. In that section we have also
mentioned that the optimal window size for the mode
estimate is $n_{pix}$=300. Both simulations and tests with real data show
that with such a number of pixels we can achieve RMS errors
on the mode smaller than 1\% for $I_{sky}\geq$100 electrons, which we
believe is a sufficient accuracy for our purposes.
For this reason, from now on we will concentrate on the specific case of
$n_{pix}$=300 and discuss several aspects of the method application.

In the previous section we have seen that, for a sufficiently large
value of $n_{pix}$, the optimal bin size can be expressed as:

\begin{equation}
\label{eq:bin}
\Delta I = 2.6 \; \frac{SNR^2_m}{N} \; \sqrt{I_{sky}}
\end{equation}

Of course this requires to know the value of $I_{sky}$, or at least to have
a rough estimate of it. For this purpose, one can approximate it with the 
median of the distribution $I_{med}$, and $SNR_m$ can be computed using 
Eq.~\ref{eq:snrm}. For $n_{pix}$=300 we have $SNR_m\simeq$150.

Once this guess value for the bin is computed, the histogram
of the intensity distribution between $I_{min}$ and $I_{max}$ is built and the
mode is found. Since in the case of skewed distributions the median
is only a rough approximation of the mode and the distribution is definitely 
not Gaussian, the actual signal-to-noise ratio $SNR^\prime_m$ in the modal 
bin is measured and from this value a new bin is recomputed as 
$\Delta I^\prime=\Delta I \; (SNR_m/SNR^\prime_m)^2$. The histogram is
recalculated and a new mode value is estimated. This procedure has a clear 
effect: if the distribution is skewed (and hence less peaked), then a larger
bin size is required to achieve the same $SNR_m$.

The method has been tested using numerical simulations of uncontaminated
300$\times$300 px windows with a known sky level $I_{sky}$ on top of 
which we have added a Poissonian noise and a typical read-out noise (6 
electrons). For each simulated field, we have then computed the mode 
\mode\/ and the deviation from the input sky background value defined as 

\begin{equation}
\label{eq:deviation}
\epsilon=\frac{\langle I\rangle - I_{sky}}{I_{sky}}
\end{equation}

\begin{figure}
\resizebox{\hsize}{!}{\includegraphics{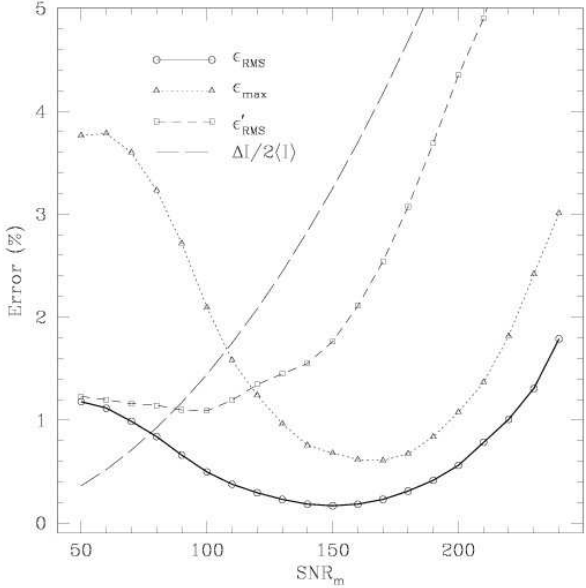}}
\caption{\label{fig:uncont} Estimated RMS mode errors (solid line)
from simulated 300$\times$300 px un-contaminated windows as a function of 
$SNR_m$. The maximum error is traced with a dotted line, while the dashed 
line indicates the RMS error when the parabolic interpolation is not used. 
Finally, the long-dashed line indicates the percentage bin half width.
The values for each $SNR_m$ level are the result of 5000 simulations.}
\end{figure}

In Fig.~\ref{fig:uncont} we present the results we have obtained 
for a very low sky background (100 electrons). The RMS deviation 
$\epsilon_{RMS}$ (solid line) reaches its minimum value (0.2\%) for $SNR_m$=150,
as expected, and so does the maximum error $\epsilon_{max}$ (dotted line).
In the same figure we have also plotted the RMS error one obtains without
using the refinement given by Eq.~\ref{eq:parabola} ($\epsilon^\prime_{RMS}$,
dashed line). This shows that the two methods give optimal results at two
different bin sizes. In such conditions the interpolation reduces the RMS 
error by about a factor of 6.
For comparison we have also plotted the bin half width (long-dashed
line), which can be assumed as an estimator of the maximum deviation when
the parabolic interpolation is not used.

On the basis of these results one would tend to adopt $SNR_m$=150, but
there is a {\em caveat} that we have to keep in mind. These results are valid
for uncontaminated distributions, where the signal-to-noise ratio
$SNR_m$ in the modal bin is reached using the bin size given by 
Eq.~\ref{eq:bin}. Since real distributions are contaminated, 
a larger bin size is required in order to achieve the same $SNR_m$ and 
this in turn generates larger errors. The simulations and extensive tests 
with real data have shown that a good compromise is reached using $SNR_m$=120 
which, for mildly contaminated distributions gives maximum errors smaller 
than 1\% for a background level of 100 electrons (see Tab.~\ref{tab:uncont}).

\begin{figure}
\resizebox{\hsize}{!}{\includegraphics{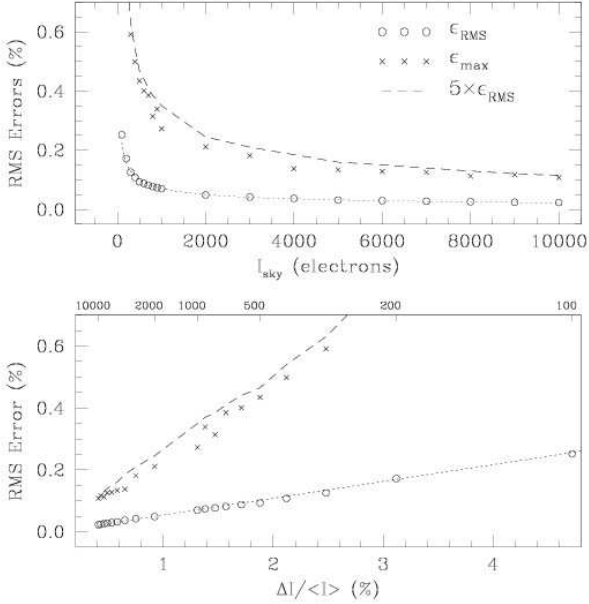}}
\caption{\label{fig:deviations} Relative errors on the mode estimate as a 
function of the sky background (upper panel) and the relative bin 
size (lower panel). The circles represent the measured RMS deviations, 
while the dotted line indicates the expected RMS (see Eq.~\ref{eq:errorlaw}).
In both panels, the crosses indicate the maximum deviation encountered 
during the simulations, while the dashed line traces the 5$\sigma$
level. Simulations were performed using $SNR_m$=120 and RON=6 electrons.
}
\end{figure}

To see how the error behaves as a function of the sky background, we
have performed another set of similar simulations, where we have
adopted $SNR_m$=120 and we have varied the input sky level 
from 10$^2$ to 10$^4$ electrons. A total of 5000 simulations per
level were executed. The results are presented in  Fig.~\ref{fig:deviations}
and summarized in Tab.~\ref{tab:uncont}.

\begin{table*}
\centering
\caption{\label{tab:uncont} Estimated errors from numerical simulations
of 300$\times$300 px uncontaminated windows for three different values of
the sky background. Maximum errors are indicated within parenthesis. A total
of 5000 simulations per sky level and $SNR_m$ value were performed.}
\begin{tabular}{cccccccccc}
\hline\hline
$SNR_m$  & f (\%) & $n_b(\pm\sigma)$ & 
\multicolumn{3}{c}{$\Delta I$/\mode  (\%)} & & \multicolumn{3}{c}{RMS error (\%) } \\
\cline{4-6} \cline {8-10}
    &    &  & $10^2$&$10^3$&$10^4$& &$10^2$&$10^3$&$10^4$ \\
\hline
120 & 16.0 &  5 &  4.7 & 1.3 & 0.4  & & 0.25 (1.00) & 0.07 (0.27)& 0.02 (0.11)\\ 
150 & 25.0 &  3 &  7.5 & 2.1 & 0.7  & & 0.17 (0.60) & 0.04 (0.16)& 0.02 (0.06)\\
\hline
\end{tabular}
\end{table*}

In the upper panel of Fig.~\ref{fig:deviations} we have plotted the percentage 
errors $\epsilon$ as a function of the background level. As expected, the 
percentage RMS error, traced with a solid line, decreases for increasing 
values of \mode, following very well the usual law for the standard error of 
the average:

\begin{equation}
\label{eq:sigma}
\sigma_{\langle I\rangle} = \sqrt{ \frac{\langle I\rangle}{N_{eff}} }
\end{equation}

with the difference that $N_{eff}$ is smaller than the total number 
$N$ of pixels. For this reason, $N_{eff}$ can be considered as the effective 
number of data points drawn from the original set one would have to use to 
get the same RMS error when adopting the average as background 
estimator\footnote{Of course this is true only if the distribution is not
contaminated, because otherwise the average gives much larger systematic
errors}. As we have seen at the end of Sec.~\ref{sec:obt}, the RMS error 
scales as $n_{pix}^{-0.75}$, and hence we have $N_{eff}=N^{0.75}$ when the 
optimal $SNR_m$ is adopted. Since we have chosen to use $SNR_m$=120, we 
expect $N_{eff}$ to be even smaller. In fact, fitting Eq.~\ref{eq:sigma} to 
the simulated data, we get $N_{eff}\sim1970$. Combining Eq.~\ref{eq:bin} and 
Eq.~\ref{eq:sigma} one gets the following relation between the bin size 
$\Delta I$ and the expected RMS error $\sigma_{\langle I\rangle}$ on the mode:

\begin{equation}
\label{eq:errorlaw}
\sigma_{\langle I\rangle} =
\frac{N}{\sqrt{N_{eff}}} \; \frac{\Delta I}{2.6\;SNR_m^2}
\end{equation}

Substituting the proper values in this expression, we obtain simply
$\sigma_{\langle I\rangle} \simeq 0.05 \; \Delta I$, which means that
for $SNR_m$=120 and moderately contaminated distributions, the expected 
RMS error on the mode is of the order of 5\% of the bin size. This is
clearly shown in the lower panel of Fig.~\ref{fig:deviations}, where we have
plotted the measured error as a function of the relative bin size derived
from the same set of simulations previously described. The match between the
expected RMS (dotted line) and the observed RMS (dashed line) is fairly good.

In both panels of Fig.~\ref{fig:deviations} we have plotted the maximum errors
encountered during the simulations. As one can see, they are well confined
within the 5$\sigma$ level (dashed line), which was computed using the
RMS error given by the simulations. More precisely, maximum errors lie
with a good approximation on the 4$\sigma$ level. From Fig.~\ref{fig:deviations}
we can conclude that the expected RMS errors on the mode estimate are below
0.3\% at all sky background levels larger than 100 electrons, while maximum
errors are always smaller than 1\%. 

The introduction of artificial stars has the effect of systematically 
increasing the mode of the distribution with respect to the real value of 
the sky background (see  Sec.~\ref{sec:bgr}). For this reason, in the
general case of contaminated distributions, we can talk about two different 
errors. While the former is a random measurement error intrinsic to the 
adopted method and to the quality of the data, the latter is a systematic 
error which depends on the intensity distribution of the contaminating 
objects. As we have said in Sec.~\ref{sec:bgr}, we want to use only the 
cases where the systematic error is smaller than some threshold value. The 
approach to this problem is described in Sec.~\ref{sec:test}, while here we 
focus only on the random errors related to the way the mode is estimated.

To evaluate the effect of contamination on the method error, we have 
performed several sets of simulations injecting a fixed number
of stars $N_*$ with random positions on a sky background of intensity
$I_{sky}$. The stars intensities $I_*$ were uniformly generated in the range 
$0 < I_* \leq (I_{sat}-I_{sky})$, where $I_{sat}$ is the detector's numerical 
saturation level ($I_{sat}\sim$ 106,000 electrons for the FORS1 detector,
high gain). Finally, a Poissonian noise was added to the artificial 
300$\times$300 px frames to simulate the photon shot noise and the mode was 
measured with the OBT using $SNR_m$=120.
Numerical tests using a more realistic intensity distribution, drawn 
from observed star counts, show that very similar contamination effects are
achieved. The only difference is that one needs to generate much more 
artificial stars in the non-uniform case, and this makes it numerically
less efficient.

As expected, the simulations show that the systematic error grows with $N_*$, 
while the random error keeps obeying to Eq.~\ref{eq:errorlaw} with a good
approximation, at least for $N_*<$200 and $\langle I \rangle \geq$ 500 
electrons. More precisely, in the case of $N_*=200$, the simulations
give $\sigma_{\langle I\rangle} \simeq 0.08 \; \Delta I$ in the range
10$^2 \leq \langle I \rangle \leq$ 10$^4$ electrons. We can safely conclude 
that the RMS errors introduced by the mode determination method we have 
described in this section are smaller than 0.3\% for $\langle I \rangle 
\geq$ 5$\times$10$^2$-10$^4$ electrons. 
Finally, the RMS error can be conservatively assumed to be 8\% of the bin 
size $\Delta I$ in the same intensity range, which corresponds to 
$N_{eff}\sim$1000. 
These values have been used in the remaining sections of this work and for
all sky brightness measurements discussed in Patat (\cite{patat}).

\section{\label{sec:delta} The $\Delta$-test}

Once the mode \mode\/ of the sky background distribution and the 
corresponding error $\sigma_{\langle I \rangle}$ have been computed, one is
left with the task of recognizing and rejecting contaminated sub-windows,
which is the topic of this section.

As we have already mentioned, in an uncontaminated image the intensity 
distribution $f(I)$ obeys the photon shot noise distribution and hence 
the expected RMS noise is given by $\sigma_p=\sqrt{\langle I\rangle}$. 
Since the left wing of $f(I)$ is much less disturbed than the right wing 
(cfr. Fig.~\ref{fig:simul}), we can estimate the underlying background noise
using only the $N_l$ pixels for which $I\leq$ \mode. As a robust estimator
we have chosen to use the Median Absolute Deviation ($MAD$) from the mode
\mode\/ applied to the left wing of the distribution:

\begin{equation}
\label{eq:mad}
MAD_l=median \{\mid I-\langle I \rangle \mid I \leq \langle I \rangle \}
\end{equation} 

We note that the use of the standard deviation instead of MAD has the
effect of over-estimating the noise in the real cases. In fact, the lower tail
of the intensity distribution is always affected by defects like bad
pixels and possible vignetting. While this fact has a rather strong influence 
on the standard deviation, it leaves the MAD unperturbed, which is much less
sensitive to the presence of outliers.

\begin{figure}
\resizebox{\hsize}{!}{\includegraphics{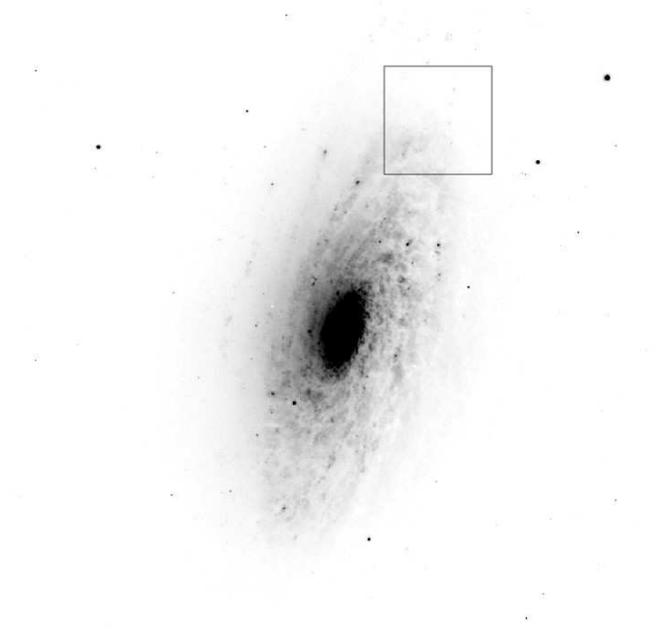}}
\caption{\label{fig:ngc3521} An image of NGC~3521 obtained by FORS1
on 04-04-2000 (R, 30 seconds, standard resolution collimator). The box 
indicates the test sub-window used for the example described in the text. 
It includes 300 $\times$ 300 pixels, which correspond to 
1$^\prime$$\times$1$^\prime$ on the sky.
}
\end{figure}

\begin{figure}
\resizebox{\hsize}{!}{\includegraphics{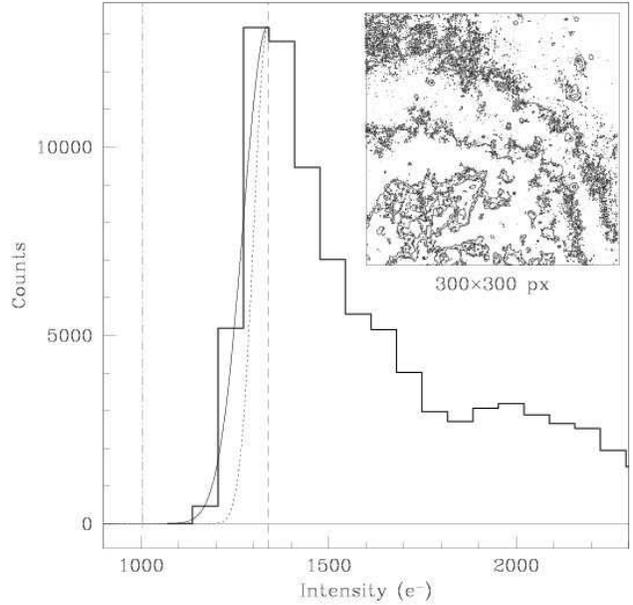}}
\caption{\label{fig:example} Intensity distribution for the sub-window
placed on a outer region of NGC~3521 (see Fig.~\ref{fig:ngc3521}). The 
vertical lines are placed at the distribution mode (dashed) and the real sky
background (dashed-dotted). The latter was manually measured in the upper 
left corner of the input image, in an object-free area. The solid and
dotted lines plotted on the left side of the distribution are two
Gaussians with $\sigma=\sigma_p$ and $\sigma=\sigma_l$ respectively.
The figure insert shows an intensity contour plot of the region. 
}
\end{figure}

One can easily show that for a Gaussian distribution the ratio between the 
standard deviation and the $MAD$ is $\sim$1.483 (Huber 1981). Hence, to have 
an estimator which has the same meaning of the standard deviation in the 
normal case\footnote{In fact, since the central value \mode\/ of the intensity
distributions we are dealing with is always larger than several hundred 
counts, the noise Poissonian distribution can be very well approximated with
a Gaussian with $\sigma=\sqrt{\langle I \rangle}$.}, we define 
$\sigma_l=1.483\;MAD_l$.
This allows one to compare directly $\sigma_p$ and $\sigma_l$ which, for an
uncontaminated distribution, should be approximately the same.
To quantify possible deviations from the Poissonian behaviour, we introduce 
the parameter $\Delta$:

\begin{equation}
\label{eq:delta}
\Delta = \frac{\sqrt{\sigma_l^2-RON^2}-\sigma_p}{\sigma_p}
\end{equation}
 
which equals 0 in the Poissonian case, whilst it gets larger and larger as the
deviation from the Poissonian distribution grows.
The idea is to try to estimate the unknown systematic error from $\Delta$,
which is a measurable parameter. To illustrate this concept we use a real 
example, where we have performed the analysis on a 300$\times$300 px 
sub-window centred on the outer parts of the large spiral galaxy NGC~3521,
as shown in Fig.~\ref{fig:ngc3521}. The corresponding intensity distribution 
is shown in Fig.~\ref{fig:example}, where the mode (indicated
by the vertical dashed line) was computed with the Optimal Binning Technique
(OBT) we have described in Sec.~\ref{sec:mode}.

\begin{figure}
\resizebox{\hsize}{!}{\includegraphics{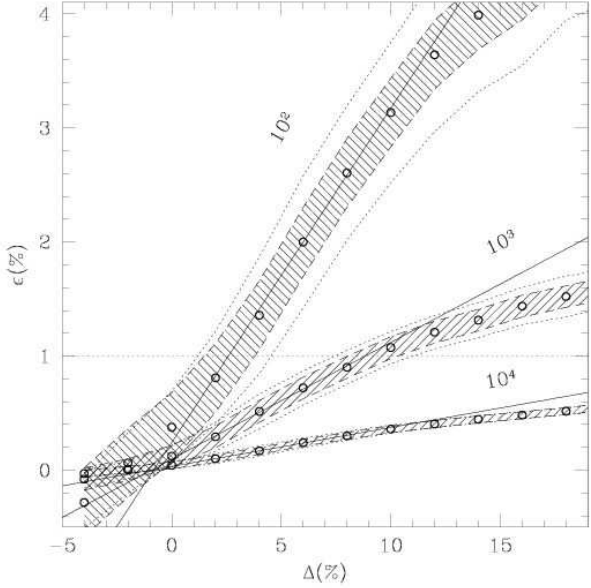}}
\caption{\label{fig:deltadev} Errors on the estimated sky background $I_{sky}$ as
a function of $\Delta$ (see text) from numerical simulations for three 
different sky background levels (10$^2$, 10$^3$ and 10$^4$ electrons). 
The continuous lines represent linear least squares fits in the range 
$0 \leq \Delta \leq 10$, while the circles are the average (systematic) 
errors. Each point is the result of 5000 simulations.
The dashed and dotted lines indicate the RMS random errors computed
from the simulations and Eq.~\ref{eq:sigma} ($N_{eff}=1000$) respectively.
The horizontal dotted line is placed at $\epsilon_{max}$=1\%.
}
\end{figure}

A superficial inspection of the input image shows already that this 
sub-window is strongly contaminated by diffuse emission from the galaxy's 
spiral arms.
The mode of the intensity distribution is in fact $\sim$34\% higher than
the background level manually measured on an object-free area of the same
image close to the upper left corner. However, the presence of strong
fluctuations within the sub-windows produces a significant increase in
the distribution core width, which becomes $\sim$80\% larger than
the expected photon shot noise ($\Delta$=0.78). This suggests that $\Delta$
may provide a tool to estimate the systematic error induced by contaminating 
sources and this, coupled with some threshold criterion, would allow us to 
recognise and disregard the critical cases.
To study this possibility, we have executed a series of simulations of 
contaminated distributions of the same type of those described in 
Sec.~\ref{sec:realcase} for several values of the input sky background 
intensity $I_{sky}$. For each
simulation one can then compute the error $\epsilon$ of the mode estimate 
(see Eq.~\ref{eq:deviation}) and measure $\Delta$. The resulting range in the 
$\Delta$ parameter can be binned to some suitable value and the 
corresponding average error $\langle\epsilon\rangle$ computed. In
Fig.~\ref{fig:deltadev} 
we have plotted the results one obtains for three different values of the 
sky background, i.e. 10$^2$, 10$^3$ and 10$^4$ electrons. The circles
represent $\langle\epsilon \rangle$, while the dashed and dotted lines 
indicate two different estimates of the random error 
$\sigma_{\langle I \rangle}$; one is the RMS deviation
directly computed from the simulated data and the other is estimated
using Eq.~\ref{eq:sigma}. While the systematic error is clearly due to
the presence of contaminating objects, the random error 
$\sigma_{\langle I \rangle}$ is related to the accuracy of the method one 
adopts to estimate the mode $\langle I \rangle$ (see Sec.~\ref{sec:mode}). 

A clear result emerges from  Fig.~\ref{fig:deltadev}: the $\Delta$ parameter
is effective in estimating the systematic error and hence gives a
statistically significant criterion to decide whether a sub-window can
be considered as unperturbed or not. 
In order to have a simple description of the behaviour of 
$\langle\epsilon \rangle$ as a 
function of $\Delta$, we have performed a linear fitting of the simulated
results in the range $0\leq\Delta \leq 10$ (see Fig.~\ref{fig:deltadev},
continuous lines). Actually $\langle\epsilon \rangle$ tends to bend for 
larger values
of $\Delta$ and this has the effect that our linear fitting slightly
overestimates the value of $\langle\epsilon \rangle$. This is not a problem, 
since
in this sense the criterion we are going to establish is just more 
restrictive. As expected, the slope of the 
$\Delta,\langle\epsilon\rangle$ 
relation is inversely proportional to the signal-to-noise ratio of the sky 
background. The best fit gives the following result:

\begin{equation}
\label{eq:syst}
\langle\epsilon\rangle = \frac{3.3}{\sqrt{I_{sky}}} \; (\Delta + 0.96)
\end{equation} 

where $I_{sky}$ is given in electrons and $\Delta,\epsilon$ are expressed
as percentage values. The fact that for 
$\Delta$=0 $\langle\epsilon\rangle\neq$0
is due to the following effect. When the sky background is contaminated
by faint stars, whose maximum intensity is comparable to
$\sqrt{I_{sky}}$, the mode and the width of the intensity distribution 
increase in such a way that $\Delta$ tends to be small, while the effective 
error is not.

We can now use the total maximum error $\epsilon_{max}$ to fix a limit 
on $\Delta$ below which we can consider a given intensity 
distributions as practically unperturbed.
Since the distribution of random errors around the systematic
deviation is Gaussian, we can conservatively assume 
$\epsilon_{max}\simeq \mid \langle\epsilon \rangle\mid 
+3 \sigma_{\langle I \rangle}$. Using Eq.~\ref{eq:syst} for $\Delta$ one gets

\begin{equation}
\label{eq:deltamax}
\Delta_{max} =  0.9 \; \sqrt{\langle I\rangle} \;
\left [ \frac{\epsilon_{max}}{3} - \sigma_{\langle I \rangle} \right ] -0.96
\end{equation}

where $\sigma_{\langle I \rangle}$ is given by the adopted mode estimator.
As it is shown in Sec.~\ref{sec:mode}, this can be approximated by 
Eq.~\ref{eq:sigma} with $N_{eff}$=1000, which gives the following expression:

\begin{equation}
\label{eq:deltamax2}
\Delta_{max} =  \frac{\sqrt{\langle I\rangle}}{3.3} \; \epsilon_{max} \; -1.9
\end{equation}

The critical value for $\Delta$ depends of course on the accuracy one wants
to reach in the estimate of the background intensity. For a typical value of 
$\epsilon_{max}$=1\%, $\Delta_{max}$ is 1.1, 7.7 and 28.4\% for sky levels of 
10$^2$, 10$^3$ and 10$^4$ electrons respectively.

Once all sub-windows in an input image have been analysed  with 
the criterion we have just discussed, a first sky brightness guess
can be obtained using the median of all selected values. This has the effect
of excluding cases like those produced by occulting masks inserted in the
focal plane. In fact, to avoid strong saturation effects, FORS instruments
allow the user to place a number of movable blades on
specific positions of the focal plane. The height of these blades is about 
20$^{\prime\prime}$, which correspond to 100 px when the SR collimator is 
used. When the occulted regions are larger
than the adopted sub-window size, there is a chance that such areas pass
the $\Delta$-test. Since the counts in those regions are very low, the
median always removes them successfully, provided that the occulted fraction 
of the field of view is not larger than 50\%, a condition which is always
fulfilled.

After this selection is done, the only background fluctuations which are 
expected to be left in the remaining sub-windows are due to a non perfect 
flat-fielding. In fact, as we have mentioned in the introduction, the use
of twilight flats introduces large scale gradients, which produce maximum
peak-to-peak deviations of 6\% from perfect flatness (see also next section).
For this reason, we have decided to operate a further refinement, choosing 
only those $n_g$ sub-windows which deviate less than 3\% from the median 
value. The final estimate of the background intensity $I_{sky}$ is eventually 
obtained computing the weighted mean 
$I_{sky}=\sum_{j=1}^{n_g} \langle I\rangle_j \; w_j / \sum_{j=1}^{n_g}w_j$,
($w_j=1/\sigma^2_{\langle I\rangle}$).
To allow for a statistically significant result, we have imposed that
$n_g\geq$5 in our automatic procedures. If this condition is not
met, then the input frame is considered as unsuitable for sky
background measures and rejected. This has the
effect of operating a first rough filtering on the input data. To be 
conservative, $n_g$ is logged together with the other relevant parameters, 
and a further and more restrictive selection is always possible.

Intensive tests with real data, as discussed in the next section, have
shown that the set of selection criteria we have discussed make the 
method reliable, robust and suitable to be implemented in a fully automatic
procedure.

\section{\label{sec:test} Testing the method on real data}

\begin{figure}
\resizebox{\hsize}{!}{\includegraphics{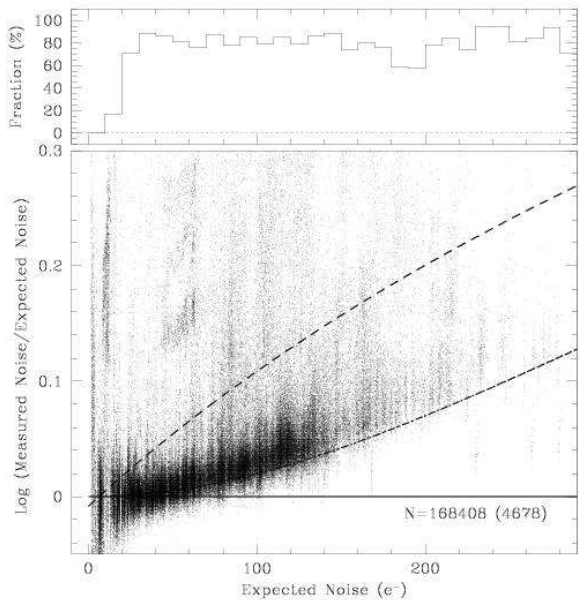}}
\caption{\label{fig:skytest} Comparison between the expected and measured
noises on a sample of more than 4600 FORS1 images. Each point
refers to a single sub-window. The solid line indicates the locus where 
the expected noise equals the measured one, while the dashed line traces the 
limit set by Eq.~\ref{eq:deltamax2}. The dashed-dotted line indicates the
expected global noise generated by the photon statistics and the flat-fielding
correction for a typical FORS1 configuration (see text for more details).
The upper panel shows the fraction of 
sub-windows which passed the $\Delta$-test as a function of the expected 
noise.
}
\end{figure}

The method we have just outlined has been tested on a sample of 
4678 FORS1 reduced frames, obtained with different filters between April 1, 
2000 and September 30, 2001. For each input frame
the results relative to all 36 sub-windows have been logged and this has
allowed us to build-up a sample with more than 168,000 entries, each
of which has been flagged according to the results of the $\Delta$-test.
The basic results are shown in Fig.~\ref{fig:skytest}, where we have plotted
the measured noise (corrected for the read-out noise) as a function of the 
expected Poissonian noise. The solid line traces the locus where the
two noises have the same value, while the dashed one indicates the limit on
the measured noise imposed by Eq.~\ref{eq:deltamax2}. All sub-windows
lying below the latter line would be selected for sky background
estimates.

The fraction of selected windows is plotted in the upper panel of the
same figure, again as a function of the expected noise. As one can see,
for noise values of less than 20 electrons (or background intensities 
smaller than $\sim$400 electrons), this fraction is very low. For larger
intensities, it reaches a roughly constant value of 80\%. This means that 
basically all frames with $\langle I\rangle\leq400$ electrons will be 
rejected.
In the adopted test sample, this however accounts for 11\%
of total number of frames only. From the test sample we can conclude 
that, on average, 86.5\% of FORS1 frames are suitable for sky-background 
measurements, 95\% of which have $n_g\geq$12.

An interesting feature visible in Fig.~\ref{fig:skytest} is the systematic
trend shown by the minimum measured noise. In fact, this tends to deviate more
and more from the expected Poissonian noise for background values
larger than $\sim$10$^4$ electrons.
This effect is explained by the following considerations. FORS1 master sky
flat-fields are usually obtained by the combination of $N_F$=3
twilight sky flats, which have typical exposure levels of 
$I_F\sim$3.5$\times$10$^4$ electrons. The resulting noise on the combined
frame is usually negligible with respect to the noise present in the
science frame to be corrected. When the background exposure level reaches high
values, this is no longer true and the noise added by the flat-fielding 
process becomes significant. The expected global noise is in fact given by 
$\sigma^2 = \sigma_p^2 + I_{sky}^2/(N_F \; I_F)$, where $I_{sky}$ is the 
background level in the input science frame. This law gives a fair 
reproduction of the observed behaviour, as it is shown in 
Fig.~\ref{fig:skytest} (dashed-dotted line), where we have used the typical 
values of $I_F$ and $N_F$ quoted above.
The few data points lying below this line are generated by observations 
obtained with the low gain mode ($\sim$0.3 ADU electron$^{-1}$), for which 
$I_F$ is about a factor of 2 larger than in the high gain mode 
($\sim$0.6 ADU electron$^{-1}$) , which is also the standard for FORS1 
imaging.
In principle, when computing $\Delta$, one could correct the measured 
noise for the flat-fielding effect. In practice,
at an exposure level of 5$\times$10$^4$ electrons the correction 
on the measured noise is of the order of 20\% and hence a very 
small number of sub-windows which are rejected by the $\Delta$ 
criterion would move to the safe region (see Fig.~\ref{fig:skytest}). 
Furthermore, 90\% of all sub-windows included in the test sample have a
background intensity smaller than 2$\times$10$^4$ electrons. At this level, 
the correction amounts to about 10\% only. For this reason we have decided to
ignore the flat-fielding effect when evaluating the deviation from the pure
Poissonian noise.

\begin{figure}
\resizebox{\hsize}{!}{\includegraphics{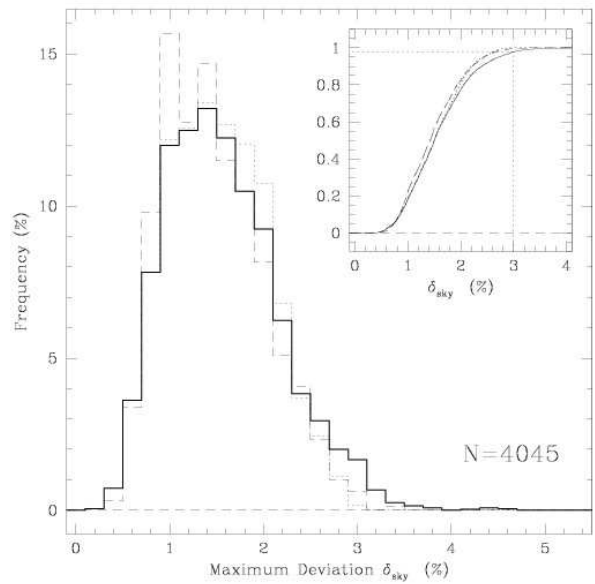}}
\caption{\label{fig:skyerrors} Maximum deviations from the sky background
weighted mean in all frames which passed the $\Delta$-test. Solid line
indicates the data from all sub-windows, the dotted line refers to
the selected windows and the dashed line corresponds to the cases where
$n_g$=36. The upper right insert shows the cumulative functions with
the same line codings.
}
\end{figure}

As we have mentioned (see also Sec.~\ref{sec:mode} for more details), the 
maximum formal errors on the mode estimate within the sub-windows are always 
smaller than 1\%. Therefore it is clear that the major contribution to the 
uncertainty on the final background estimate is due to the non perfect 
flat-fielding, which introduces smooth large scale gradients in the
reduced images. Due to the systematic nature of the effect, it does not make 
any statistical sense to adopt the formal error on the weighted mean to 
estimate this uncertainty and for this reason we have preferred to use
the maximum deviation $\delta_{sky}$ measured in the $n_g$ selected windows.
Of course, when $n_g\ll$36 and all the accepted sub-windows are 
concentrated in a portion of the frame, the use of $\delta_{sky}$ can lead
to an error underestimate. In the case of our test sample, this effect is
present for $n_g \lesssim $10, where the estimated error is a factor of two
smaller than for larger $n_g$ values. Since the large majority of our data
had $n_g>$12, this does not affect significantly our error estimates. This
effect can be anyway reduced adopting larger values for the minimum
number of sub-windows that must survive the $\Delta$-test.

The results produced applying our method to the test sample are shown
in Fig.~\ref{fig:skyerrors}, where we present the distribution
of $\delta_{sky}$ derived from the 4045 images which passed the $\Delta$-test.
The solid line refers to the results one obtains using all windows which
passed the $\Delta$-test, while the dotted line corresponds to the
values obtained from the $n_g$ selected windows only. While the latter
by definition drops to 0 at $\delta_{sky}$=3\%, the former shows also the
deviating cases at larger $\delta_{sky}$ which are, however, less than
3\% of total. We emphasise that 20\% of the measurements
have $\delta_{sky}\leq$1\%, while this fraction grows to 77.5\% for
$\delta_{sky}\leq$2\%. The median value in the whole sample is 1.5\%, which
can be regarded as the typical maximum error in our measurements.
It is finally interesting to note that the maximum error distribution
one obtains using only those cases where all sub-windows are used for the
final estimate ($n_g$=36, 1594 images, 39.5\% of total) is not very different
from the one which corresponds to the general case (Fig.~\ref{fig:skyerrors}, 
dashed line). Since the images for which all 36
sub-windows are selected are {\it bona fide} not affected by significant
contamination, this confirms that the $\delta_{sky}$ distribution we observe
is really due to large scale gradients and not to contaminated sub-windows
which escaped the $\Delta$-test.

\begin{figure}
\resizebox{\hsize}{!}{\includegraphics{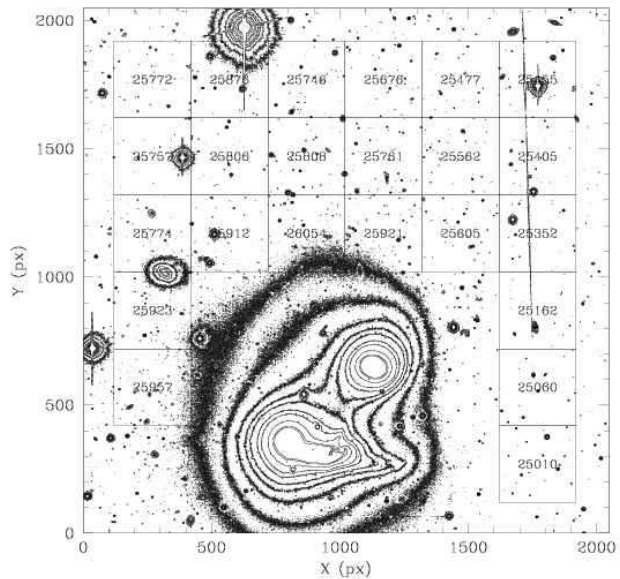}}
\caption{\label{fig:interact} An example of a critical case: a group of
interacting galaxies. The original 660 seconds image was taken in the 
R band on 03-05-2000, using the SR collimator (0\as2 pixel$^{-1}$). The 
lowest contour was set at a 5 sigma level above the minimum sky background 
(measured in the upper right corner). Marked boxes indicate the sub-windows
which were automatically judged as suitable for sky background 
determination according to our method; the number in each box shows the 
estimated sky background in electrons. The strip on the right side is a
satellite trail.
}
\end{figure}

The efficiency of the method in recognising critical cases has been checked 
directly, with a visual inspection of a large number of cases included in the
test sample. The conclusion is that the method is reliable and does not lead
to artificial over-estimates of the sky background.
As an example of these capabilities, in Fig.~\ref{fig:interact} we present 
a critical case drawn from our data sample: an $R$-band image of interacting
galaxies. The boxes indicate the sub-windows 
which have passed the $\Delta$-test and which would be used for the first 
estimate.
The lowest contour was traced at the 5 sigma level of the sky background
(manually measured on a star-free area in the right upper corner). All
selected sub-windows lie outside this contour, which reasonably defines
the region where the galaxies certainly contribute to the background.
As a matter of fact, there is a gradient within the accepted 
sub-windows, but the peak-to-peak difference is about 3\% only, a value 
which is still consistent with the flat-fielding accuracy. 
As expected, all critical regions are disregarded. The final sky brightness 
is computed using the weighted mean on 23 sub-windows. It is interesting 
to note that even though some sub-windows include outer parts of the 
galaxies, the background value computed using our algorithm is fully 
consistent with that of visually selected object-free regions.

\section{\label{sec:conclusion}Conclusions}

In this paper we have presented a numerical algorithm to
estimate the sky background in CCD imaging data. Due to the practical purpose
this technique was designed for, we have optimized its implementation
for FORS1; however, the algorithm is based on very general assumptions
and it can be used for any CCD imager with a sufficiently wide field of view
($\geq$ 5$^\prime \times$5$^\prime$).

The fulfilment of this requirement, coupled to the large size of the 
detectors currently available, allows one to estimate the mode of the image 
intensity distribution directly from its histogram, with a typical accuracy 
of 1\% or better (Sec.~\ref{sec:mode}).
In order to identify suitable regions within the image, the analysis is
performed in smaller sub-windows which, in the case of FORS1, have a
size of 1$^\prime \times$1$^\prime$ and include 9$\times$10$^4$ pixels. The 
possible presence of contaminating objects within these areas is detected 
studying the shape of the intensity distribution.

The criterion to reject such regions from the final sky background estimate,
that we have indicated as $\Delta$-test, has been established via numerical 
simulations and it is based on the effects that perturbing objects have on 
the left wing of the local intensity distribution (Sec.~\ref{sec:delta}).

The method has been designed to be robust and reliable under a large variety 
of conditions. The tests have shown that it can be safely used in fully 
automatic procedures (Sec.~\ref{sec:test}) and therefore it is suitable for 
processing large data volumes. The tests on real images have also shown that 
the final accuracy is determined mostly by the flat-fielding quality on large
scales. In the specific case of FORS1 this is typically of the order of 
2$-$3\% (peak-to-peak) across the whole field of view. This sets a lower 
limit for the study of night sky brightness variations on the arcminutes 
scale, since they cannot be disentangled from those artificially induced by 
the flat-fielding process.

As far as the contribution of faint stars to the sky background is
concerned, the simulations show that our algorithm is undisturbed by the 
presence of stellar objects with peak intensity $I_*\geq5\sigma_p$. In the
case of sky background dominated images, the magnitude of those sources is 
given by the following expression:

\begin{equation}
\label{eq:faint}
m_f=m_0-2.5\;log \left [5.7\;\frac{FWHM^2}{p^2}\right ] -
1.25\;log \left [ \frac{r_{sky}}{t}\right ]
\end{equation}

where $m_0$ is the photometric zero point in a given passband, $p$ is the
pixel scale (in arcsec px$^{-1}$), $r_{sky}$ is the sky background rate
(in e$^-$ s$^{-1}$), $t$ is the exposure time (in seconds) and $FWHM$ is
the seeing (in arcsec). For example, FORS1 $V$ frames obtained through the
SR collimator ($p$=0.2 arcsec px$^{-1}$) become sky background dominated in 
about 25 seconds (see Patat \cite{patat}). With this exposure time,
a seeing of 1$^{\prime\prime}$ and the typical FORS1 zero point in the
$V$ passband ($m_0\simeq$28), Eq.~\ref{eq:faint} gives $m_f\sim$22.8,
which is about 10 magnitudes fainter than the value for typical
photoelectric sky brightness surveys (Walker \cite{walker88}). Such
faint objects contribute to less than 1\% to the total brightness
(Roach \& Gordon \cite{roachgordon}) and therefore we can conclude that
our method is practically free from being biased by the inclusion of
faint foreground point sources.

Finally, to asses the speed performance of our algorithm, we have compiled and
executed a C coded version on a moderately fast Linux PC (Penthium III 500 
MHz, 256 MB RAM). On such a machine, the analysis of a 2048$\times$2048 px 
image requires less than 6 seconds, making it suitable for on-line processing.

The method we have presented here has been extensively used in the
ESO-Paranal night sky brightness survey, which made use of more than 3900
$UBVRI$ FORS1 frames collected from April 2000 to September 2001. The
results are presented and discussed in Patat (\cite{patat}).

\begin{acknowledgements}

I am profoundly indebted to Martino Romaniello, for the illuminating 
discussions, useful advises and stimulating suggestions. I also would like 
to express my gratitude to Bruno Leibundgut, Dave Silva, Gero Rupprecht
and Jean-Gabriel Cuby for carefully reading the original manuscript.
Finally, I wish to thank the referee, H. Hensberge, for his
comments and suggestions, which greatly improved the quality of the paper.

\end{acknowledgements}

\end{document}